\documentclass[journal]{IEEEtran}

\usepackage{graphicx}
\usepackage{multicol}
\usepackage{amssymb}
\usepackage{amsmath}
\usepackage{booktabs,multirow}
\usepackage{color}
\usepackage{url}
\usepackage{cite}
\usepackage{textcomp}
\usepackage{algorithm}
\usepackage{algorithmic}

\newcommand{\tA}{\mathcal{A}}

\newcommand{\tD}{\mathcal{D}}

\newcommand{\di}{\mathrm{diag}}
\newcommand{\tG}{\mathcal{G}}
\newcommand{\tV}{\mathcal{V}}

\newcommand{\tL}{\mathcal{L}}

\newcommand{\tN}{\mathcal{N}}

\newcommand{\bT}{\mathbb{T}}
\newcommand{\bR}{\mathbb{R}}

\newtheorem{thm}{Theorem}

\newtheorem{rem}{Remark}
\newtheorem{coro}{Corollary}

\renewcommand\IEEEkeywordsname{Keywords}

\begin{document}

\title{Cooperative Learning for P2P Energy Trading via Inverse Optimization and Interval Analysis}

\author{Dinh Hoa Nguyen,~\IEEEmembership{Member,~IEEE} 
\thanks{Dinh Hoa Nguyen is with the International Institute for Carbon-Neutral Energy Research (WPI-I$^2$CNER), and the Institute of Mathematics for Industry (IMI), Kyushu University, Fukuoka 819-0395, Japan. E-mail: $hoa.nd@i2cner.kyushu-u.ac.jp$. }
}

\maketitle

\begin{abstract}
Peer-to-peer (P2P) energy systems have recently emerged as a promising approach for integrating renewable and distributed energy resources into energy grids to reduce carbon emissions. However, market-clearing energy price and amounts, resulted from solving optimal P2P energy management problems, might not be satisfactory for peers/agents. This is because peers/agents in practice do not know how to set their cost function parameters when participating into P2P energy markets. To resolve such drawback, this paper proposes a novel approach, in which an inverse optimization problem is formulated for peers/agents to cooperatively learn to choose their objective function parameters, given their intervals of desired energy prices and amounts. The result is that peers/agents can set their objective function parameters in the intervals computed analytically from the lower and upper bounds of their energy price and amounts, if the ratio of their maximum total buying and selling energy amounts lies in a certain interval subject to be learned by them. A case study is then carried out, which validates the effectiveness of the proposed approach. 
   
\end{abstract}


%\IEEEpeerreviewmaketitle

\section*{Nomenclature}
\addcontentsline{toc}{section}{Nomenclature}
\begin{IEEEdescription}[\IEEEusemathlabelsep\IEEEsetlabelwidth{$P_{i,G}$, $P_{l,RDP}$}]
\item[MAS] Multi-agent system.
%\item[ICT] Information and communication technology.
\item[P2P] Peer to peer.
\item[DER] Distributed energy resource.
\item [$P_{ij}$, $P_{i}$] Traded power/energy between peers $i$ and $j$ and vector of peer $i$ traded powers/energy.
\item [$P_{i,tr}$] Total trading power of peer $i$ [kW].
%\item [$\underline{P}_{i,tr}$, $\overline{P}_{i,tr}$] Lower and upper bound of trading power amount of prosumer $i$ [kW].
\item [$P_{i,b}$, $P_{i,s}$] Total selling or buying power of peer $i$ [kW].
\item [$\underline{P}_{i,b}$, $\overline{P}_{i,s}$] Lower bound and upper bound of power/energy amount for selling and buying prosumer $i$, respectively [kW].
\item [$a_{i,b}$, $b_{i,b}$] Parameters of selling prosumer/peer $i$ cost function.
\item [$a_{i,s}$, $b_{i,s}$] Parameters of buying prosumer/peer $i$ cost function.
\item [$t, \bT$] Time step, and the number of considered time steps.
\item[$\tG, \, \tA, \, \tD, \, \tL$] P2P interconnection graph, its adjacency, degree, and Laplacian matrices.
\item[$\mathbf{1}_n$, $I_n$] Vector with $n$ elements equal to $1$, and $n\times n$ identity matrix.
\item[$\di\{\}$, $vec()$] Diagonal or block-diagonal matrices, and stacked vector. 
\item[$\bR$, $\bR^{n}$, $\bR^{n\times m}$] Set of real numbers, real $n$-dimension vectors, and real matrices with dimensions $n\times m$.
\end{IEEEdescription}

%%%%%%%%%%%%%%%%%%%%%%%%%%%%%%%%%%%%%%%%%%%%%%%%%%%%%%%%%%%%%%%%%%%%%%%%%%%%%%%%%%%%%%%%%%%%%%%%%%%%%%%%%%%%%%%%%%%%
\section{Introduction}

The wide adoption of renewable and DERs all over the world, as an effort to reduce carbon emissions, not only poses many challenges to the operation and management of energy systems, but also brings opportunities to develop novel approaches for revolutionizing energy systems. P2P energy system is such an approach recently attracts much attention, due to its suitability for renewable and DERs integration and many other advantages \cite{Baez-Gonzalez18,Sousa19,Tushar18,Tushar20}. For example, P2P energy system has potential for reduction of energy losses (through energy transfer in local areas), flexibility for provision of demand side management services, better security and privacy protection with distributed ledger technologies, and multiple possible business models \cite{Tushar18,MorstynP2P18,Sousa19,Moret19}, to name a few.  

In addition, P2P energy system helps strengthen the role of prosumers, who are both energy producers and consumers, to proactively participate in energy markets instead of being just passive consumers. Each peer/prosumer in a P2P energy system can directly communicate and trade energy with other peers/prosumers (similar to the P2P protocol in computer science), which is essentially different from that in pool-based energy markets. Therefore, new approaches for the operation and management of P2P energy markets need to be developed. Hitherto, a body of works has been proposed in the literature to derive distinct P2P energy trading mechanims, e.g. bilateral contracts \cite{Sorin19,Baroche19,MorstynP2P19b,Khorasany19}, game theory based \cite{Tushar19,Tushar18,Moret19,Cadre20}, distribution optimal power flow \cite{GuerreroA19}, supply-demand ratio based pricing \cite{NLiu17}, mixed performance indexes \cite{Werth18}, Lyapunov optimization \cite{NLiu18}, multi-class energy management \cite{MorstynP2P19a}, continuous double auction \cite{GuerreroA19}, etc. 
  
From the algorithm viewpoint, a P2P energy system can also be regarded as an MAS, where each peer/prosumer is cast as an agent. Accordingly, multi-agent-based optimization and control algorithms have been derived for P2P energy systems (see e.g., \cite{Sorin19,Nguyen-TPWRS20} and references therein). As a result, the communication and trading between prosumers can be made autonomous, in which each agent acts on behalf of a prosumer. %In other words, artificial intelligence, in form of MAS, helps advance the operation and management of P2P energy systems.  

Different characteristics of P2P energy markets have been explicitly shown in \cite{Nguyen-TPWRS20} including unique or multiple P2P market clearing prices, and clustered P2P energy markets, partly due to unsuccessful energy transactions. As such, two fundamental assumptions have been commonly employed in the literature of P2P energy systems research, one is the successful trading of all prosumers, and the other is the right selection of cost function parameters by each prosumer to obtain expected energy transactions. However, those assumptions can be easily violated in realistic situations because of distinct preferences on the amount of powers and their prices to be traded between prosumers, and the lack of knowledge on the relation between such energy preferences with cost function parameters. 

To cope with the challenge on relaxing such assumptions, a heuristic approach for selecting parameters of prosumers' cost functions in P2P energy systems composing of multiple selling and multiple buying prosumers has been introduced in \cite{Nguyen-TPWRS20}. This heuristic approach has been shown quite effective in assuring the success of P2P energy trading and the increase of trading energy amounts, and has been applied to several decentralized P2P energy trading systems \cite{Nguyen-TPWRS20,Nguyen-IJEPES20}. In a more recent work \cite{Nguyen-Access20}, an analytical method has been introduced for cooperative learning of prosumers/agents, but only for special cases where only one selling or one buying prosumer exists. The current work follows the above research line to formally formulate that problem in form of an inverse optimization problem, and proposes an analytical cooperative learning approach to solve it for the general case of multiple selling and multiple buying prosumers.       

The contributions of this research are as follows. 
\begin{itemize} 
	\item A decentralized cooperative learning approach for analytically choosing parameters of prosumers/agents' cost functions to guarantee successful trading with expected energy price and power/energy amounts, in P2P energy systems consisting of mutiple buying and multiple selling prosumers/agents. This is the first time such result has been presented in the literature. 
	\item Interval analysis is employed in the proposed approach, where prosumers/agents initially set their preferred intervals of trading energy price and amounts, then cooperatively learn to analytically select their cost function parameters in specific intervals to satisfy their preferences. 
\end{itemize}
It is also worth emphasizing here that even though the proposed learning approach is presented for P2P energy markets, it is certainly applicable for other systems and applications having the same formulation. 
%This clearly shows the strength of artificial intelligence, through multi-agent-based optimization, control, and learning, to solve practical problems.  

The rest of this paper is organized as follows. Section \ref{sysmod} presents the considered P2P energy systems and their inverse optimization problem. Then a decentralized cooperative learning approach for prosumers/agents is proposed in Section \ref{cooper-learning} to solve the introduced inverse optimization problem. A case study is presented in Section \ref{num} to demonstrate the  proposed approach. Finally, Section \ref{sum} concludes the paper and provides directions for future research.

%%%%%%%%%%%%%%%%%%%%%%%%%%%%%%%%%%%%%%%%%%%%%%%%%%%%%%%%%%%%%%%%%%%%%%%%%%%%%%%%%%%%%%%%%%%%%%%%%%%%%%%%%%%%%%%%%%%%
\section{P2P Electricity Trading Problem}
\label{sysmod}

Consider the P2P energy trading in an energy system consisting of $n$ prosumers during a time interval $[1,\bT]$, in which each prosumer is regarded as a peer or agent who can both produce and consume power. In addition, prosumers are assumed to behave non-strategically. 

Let $P_{ij}(t)$ be the traded power between prosumers $i$ and $j$ at time step $t$, $P_{ij}(t)>0 (<0)$ means prosumer $i$ sells to (buy from) from prosumer $j$. Furthermore, we assume that at each time step a prosumer only sells or buys power, but not to do both, to simplify the prosumers power trading.   

%=====================================
\subsection{Communication Structure Between Prosumers}

\begin{figure}[htpb!]
		\centering
		\includegraphics[scale=0.4]{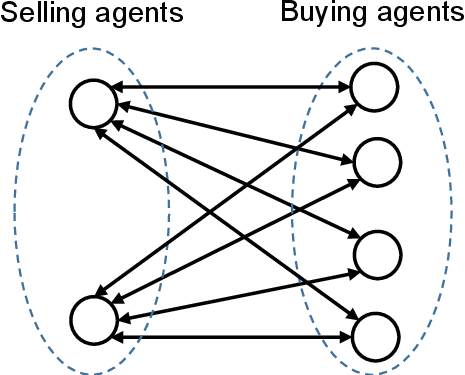}
		\caption{Illustration for the communication structure between prosumers/agents.}
		\label{com-struct}
	\end{figure}

The inter-prosumer communication structure, depicted in Figure \ref{com-struct},   is bipartite and time-varying, and is represented by a graph $\tG(t)$ which is undirected because of the bilateral trading between prosumers. The node set of $\tG(t)$ consists of two disjoint subsets corresponding to selling and buying prosumers, which are denoted by $\tV_s(t)$ and $\tV_b(t)$, respectively. 

Next, denote $\tN_i(t)$ the neighboring set of prosumer $i$ at time step $t$, i.e., the set of other communicated prosumers. 
Let $0 \leq a_{ij} \leq 1$ be elements of the inter-prosumer communication matrix $\tA(t)$, where $a_{ij}(t) > 0$ means prosumers $i$ and $j$ are connected at time step $t$, and $a_{ij}(t)=0$ otherwise. Moreover, $\tA(t)$ is a symmetric and doubly-stochastic matrix (i.e. row sums and column sums of $\tA(t)$ are all equal to 1). 
%The degree matrix $\tD(t)$ is defined by $\tD(t)=\di\{d_{i}(t)\}_{i=1,\ldots,n}$, where $d_{i}(t) \triangleq \sum_{j \in \tN_{i}(t)}{a_{ij}(t)}$. 
%Then the Laplacian matrix $\mathcal{L}(t)$ associated to $\mathcal{G}(t)$ is defined by $\tL(t)=\tD(t)-\tA(t)$.  

%=====================================
\subsection{Objective Function}

Denote $n_{i}(t) \triangleq |\tN_i(t)|$, $P_i(t) \triangleq vec(P_{ij}(t))_{j \in \tN_i(t)}$, and $P_{i,tr}(t) \triangleq \mathbf{1}_{n_{i}(t)}^TP_i(t)$. Here, the notation $P_{i,tr}(t)$ is used to simplify the representation of results, and subsequently it will be replaced by $P_{i,b}(t)$ or $P_{i,s}(t)$ depending on whether the associated prosumer/agent is a seller or a buyer to clearly distinguish its role. 

Let $C_i(P_{i}(t))$ denote the overall cost function of prosumer $i$ for trading in the P2P energy market. Individual components of $C_i(P_{i}(t))$ are presented below. %assumed to have the following forms.  
 %of peer $i$, 
\begin{subequations}
\begin{align}
\label{cost-1}
C_{i,1}(P_{i}(t)) &= a_i(t) P_{i,tr}^2(t) + \tilde{b}_i(t) P_{i,tr}(t) \\ %+ c_i(t) 
%\end{equation}
%\begin{equation}
%\label{cost-2}
%C_{i,2}(P_{i}(t)) &= \sum_{j \in \tN_i} d_{ij} P_{ij}(t) \\
%\end{equation}
%\begin{equation}
\label{cost-3}
C_{i,2}(P_{i}(t)) &= \beta P_{i,tr}(t)
\end{align}
\end{subequations} 
Eq. \eqref{cost-1} is an utility function, assumed to be quadratic, with time-varying private parameters $a_i(t)>0$ and $\tilde{b}_i(t)>0$ which show the time-varying and complex behaviors of prosumers. 
%Eq. \eqref{cost-2} is a bilateral trading cost with bilateral trade weights $d_{ij}$ utilized for the purposes of product differentiation and consumer involvement (see e.g., \cite{Sorin19}, \cite{Nguyen-TPWRS20}).  
Eq. \eqref{cost-3} is the implementation cost paid to the bulk power grid for physically executing P2P energy transactions, with fixed rate $\beta>0$. There would be another component representing the bilateral trading cost between prosumers/agents (see e.g., \cite{Sorin19}, \cite{Nguyen-TPWRS20}), however it is ignored here for simplicity. 

Denote $b_i(t) \triangleq \tilde{b}_i(t) + \beta$, then $C_i(P_{i}(t))$ is the sum of its components in \eqref{cost-1}--\eqref{cost-3}, 
\begin{equation}
\label{cost}
C_{i}(P_{i}(t)) = a_i(t) P_{i,tr}^2(t) + b_i(t) P_{i,tr}(t) %+ \sum_{j \in \tN_i} d_{ij} P_{ij}(t)
\end{equation} 
which is a convex function.  

%=====================================
\subsection{System Constraints}

The first constraint is on the bilateral trading power, i.e.,
\begin{equation}
\label{e-1}
P_{ij}(t) + P_{ji}(t) = 0  ~\forall \; j \in \tN_i(t), \ t=1,\ldots,\bT 
\end{equation}
The next constraint is on the limits of power can be traded, %as follows.  
\begin{equation}
\label{e-2}
%\underline{P}_{i,tr} \leq P_{i,tr}(t) \leq \overline{P}_{i,tr} ~\forall \; t=1,\ldots,\bT 
\underline{P}_{i,b}(t) \leq P_{i,b}(t) \leq 0; ~ 0 \leq P_{i,s}(t) \leq \overline{P}_{i,s}(t) ~\forall \; t=1,\ldots,\bT 
\end{equation}
Such limits are based on the own profiles of power generation and consumption of each prosumer, as well as the guidance from power system operators, if exists, to avoid physical problems (e.g., constraints on grid voltage and frequency) which may affect to the grid stability, reliability, etc.  
Power flow constraints are not included, and is assumed to be handled by power network operators, which is paid by prosumers as shown in the cost \eqref{cost-3}. 

%=====================================
\subsection{Forward Optimization Problem}

%Subsequently, 
The market-clearing problem for P2P energy markets, which is often studied in the literature, is described as follows.

{\bf $\blacksquare$ Forward Optimization Problem for P2P Energy Market:} Given parameters $a_i$ and $b_i$ of prosumers/agents' cost functions \eqref{cost}, find the market-clearing energy price and trading amounts of all prosumers/agents.

%Suppose that $\zeta_{ij}$ is the energy price for the energy transaction between prosumers/agents $i$ and $j$, then $\sum_{j \in \tN_i(t)} \zeta_{ij} P_{ij}(t) - C_{i}(P_{i}(t))$ depicts the monetary benefit if prosumer $i$ is a seller, whereas $\sum_{j \in \tN_i(t)} \zeta_{ij} P_{ij}(t) - C_{i}(P_{i}(t))$ reveals the satisfaction level if prosumer $i$ is a buyer. This is similar to the welfare defined in the dynamic social welfare maximization problem for the pool-based markets (see e.g., \cite{Nguyen-TSG17,Nguyen-TIE19} and references therein). Hence, the optimal clearing strategy for P2P energy trading is to maximize the welfare of all prosumers, i.e. to maximize
%\begin{align*}
%& \sum_{i \in \tV_s(t)} [ \sum_{j \in \tN_i(t)} \zeta_{ij} P_{ij}(t) - C_{i}(P_{i}(t)) ]  \\
%& + \sum_{i \in \tV_b(t)} [ \sum_{j \in \tN_i(t)} \zeta_{ij} P_{ij}(t) - C_{i}(P_{i}(t)) ] 
%\end{align*}
%Due to the fact that $\zeta_{ij}=\zeta_{ji}$ and $P_{ij}(t)+P_{ji}(t)=0$, this is equivalent to solve the following convex optimization problem.

This forward optimization problem is written as: 
\begin{subequations}
\begin{align}
\label{cost-p2p}
\min \; & \sum_{t=1}^{\bT} \sum_{i=1}^{n} C_i(P_{i}(t))    \\
\label{balance-p2p}
\textrm{s.t.} ~ & P_{ij}(t) + P_{ji}(t) = 0 ~\forall \; j \in \tN_i(t)  \\
\label{bounded-power-p2p}
~ & \underline{P}_{i,b}(t) \leq P_{i,b}(t), i \in \tV_b(t); ~ P_{i,s}(t) \leq \overline{P}_{i,s}(t), i \in \tV_s(t) \\
\label{pos-neg}
~ & P_{i,b}(t) \leq  0 ~\textrm{if} ~i \in \tV_b(t); ~P_{i,s}(t) \geq  0 ~\textrm{if} ~i \in \tV_s(t)
\end{align}
\label{p2p}
\end{subequations} 
which is in fact similar to the dynamic social welfare maximization problem for the pool-based markets (see e.g., \cite{Nguyen-TSG17,Nguyen-TIE19} and references therein). 
As shown in \cite{Nguyen-TPWRS20}, \eqref{p2p} can be solved at each time step, hence the time index will be ignored for conciseness of mathematical expressions. 
Resolving \eqref{p2p} then gives us the P2P market clearing energy price (which is equal to the dual variable associated to the equality constraint \eqref{balance-p2p}) and power trading amounts $P_{ij}$.

%=====================================
\subsection{Inverse Optimization Problem}
\label{inv-opt-prob}

An issue arises when solving the forward optimization problem \eqref{p2p} is that some prosumers might be unsuccessful in trading, as shown in \cite{Nguyen-TPWRS20}.
Another issue encountered in practical situations is that the energy trading price, i.e. the dual variable associated with the equality constraint \eqref{balance-p2p}, obtained by solving the forward optimization problem \eqref{p2p}, is not satisfied by all prosumers. Therefore, the current research aims to overcome above issues by investigating an inverse problem, as follows. 
%proposes an analytical approach to guarantee the energy trading success of all prosumers by formulating and solving an inverse optimization problem  

{\bf $\blacksquare$ Inverse Optimization Problem for P2P Energy Market:} Each participated prosumer/agent $i,i=1,\ldots,n,$ sets the following {\it a priori}. 
\begin{itemize}
	%\item Quadratic form of its cost function as in \eqref{cost}.
	\item Preferred range $\left[\underline{\lambda}_i, \overline{\lambda}_i\right]$ of P2P energy price.
	\item Preferred range $\left[\underline{P}_{i,b}, 0) \right.$ or $\left. (0, \overline{P}_{i,s}\right]$ of power amounts when it is a buying or selling prosumer; respectively.
\end{itemize}
Find the parameters $a_i$ and $b_i$ of the prosumer's cost function \eqref{cost}  to achieve successful energy trading with above desired quantities. 

In this inverse optimization problem, the solution (trading energy price and amounts of prosumers/agents) and constraint sets (bilateral trading balance and ranges of power amounts) are known, and its goal is to derive parameters of prosumers/agents' objective functions. This problem is very realistic, but is rarely studied in the literature. A recent work \cite{Nguyen-Access20} investigated it in an informal way for a special context where only one selling or one buying prosumer exists. The general scenario of multiple buying and multiple selling prosumers has not been studied hitherto. 

\begin{figure}[htpb!]
		\centering
		\includegraphics[scale=0.45]{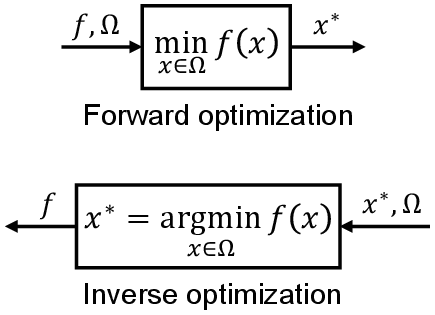}
		\caption{Principles of forward and inverse optimization problems.}
		\label{inv-opt}
	\end{figure}

General principles of the forward and inverse optimization problems are illustrated in Figure \ref{inv-opt} for clarity in their differences. 

%%%%%%%%%%%%%%%%%%%%%%%%%%%%%%%%%%%%%%%%%%%%%%%%%%%%%%%%%%%%%%%%%%%%%%%%%%%%%%%%%%%%%%%%%%%%%%%%%%%%%%%%%%%%%%%%%%%%
\section{Prosumers Cooperative Learning}
\label{cooper-learning}

Define the following Lagrangian associated to \eqref{p2p},
\begin{equation}
\label{Lagrangian}
L(P_{ij},\lambda_{ij}) = \sum_{i=1}^{n} C_i(P_{i}) - \sum_{i=1}^{n} \sum_{j \in \tN_i} \lambda_{ij}(P_{ij} + P_{ji})  
\end{equation}
where $\lambda_{ij}$ are the Lagrange multipliers associated to the power trading equations \eqref{balance-p2p}, which are regarded as the market clearing prices for energy transactions between pairs of prosumers. 

When the inequality constraints \eqref{bounded-power-p2p}--\eqref{pos-neg} are omitted and the communication graph between successfully traded peers is connected, 
the P2P energy clearing price was shown to be unique \cite{Nguyen-TPWRS20}, which is computed by 
\begin{equation}
\label{eq-price}
\lambda^{\ast} = \frac{\sum b_j/a_j}{\sum 1/a_j}
\end{equation}
And the optimal total trading power for each peer/prosumer is
\begin{equation}
\label{eq-power}
P_{i,tr}^{\ast} = \frac{1}{2a_i}(\lambda^{\ast}-b_i) = \frac{\sum b_j/a_j}{2a_i\sum 1/a_j} - \frac{b_i}{2a_i}
\end{equation}
The formulas \eqref{eq-price}--\eqref{eq-power} of the forward optimization problem will serve as basis for the proposed cooperative learning approach between prosumers/agents obtained via solving the inverse optimization problem.

%---------------------------------------
\subsection{Main Results}

First, prosumers with individual price ranges $\left[\underline{\lambda}_i, \overline{\lambda}_i\right]$ need to negotiate for obtaining the same range. 
There are several methods to do so, e.g. min-max and averaging. In the former, the maximum of lower bounds and minimum of upper bounds of prosumers' price intervals are derived, while in the latter the average of lower and upper bounds are obtained.  
For simplicity, we assume that the preferred price intervals of prosumers are overlapped, hence either of those strategies could be used. Here, we choose the averaging method, similar to that in \cite{Nguyen-Access20}, therefore details are ignored for conciseness. 
Next, denote 
\begin{align*}
\tilde{a}_{i,s} &= \frac{1}{a_{i,s}}, ~\tilde{a}_{i,b} = \frac{1}{a_{i,b}} \\
\underline{b}_{b} &= \min \{b_{i,b}\}, ~\overline{b}_{b} = \max \{b_{i,b}\} \\
\underline{b}_{s} &= \min \{b_{i,s}\}, ~\overline{b}_{s} = \max \{b_{i,s}\} 
%\underline{b}_{b} &= \min_{i \in \tV_{b}} \{b_{i,b}\}, ~\overline{b}_{b} = \max_{i \in \tV_{b}} \{b_{i,b}\}, \\
%\underline{b}_{s} &= \min_{i \in \tV_{s}} \{b_{i,s}\}, ~\overline{b}_{s} = \max_{i \in \tV_{s}} \{b_{i,s}\} 
\end{align*}
Based on interval analysis, the following theorem reveals the intervals for parameters of prosumers such that the inverse optimization problem in Section \ref{inv-opt-prob} is solved. 

\begin{thm}
\label{para-cond-1}
Having the consensus price range $[\underline{\lambda},\overline{\lambda}]$, the lower bound $\underline{P}_{i,b}$ of power amount to be bought by buying prosumers, and the upper bounds $\overline{P}_{i,s}$ of power amounts to be sold by selling prosumers, the following conditions are sufficient to strictly satisfy the constraints \eqref{bounded-power-p2p}--\eqref{pos-neg},
\begin{subequations}
\label{ab-1}
\begin{align}
\label{b-selection-1}
\underline{\lambda}\, &\leq \,\underline{b}_{s}\, \leq \,\overline{b}_{s}\, < \,\underline{b}_{b}\, \leq \,\overline{b}_{b}\, \leq \,\overline{\lambda}  \\
\label{a-cond-s-1}
\tilde{a}_{i,b} &< \frac{-2\underline{P}_{i,b}}{\overline{b}_{b}-\overline{b}_{s}} \, \Leftrightarrow \, a_{i,b} > \frac{\overline{b}_{b}-\overline{b}_{s}}{-2\underline{P}_{i,b}} \\
\label{a-cond-b-1}
\tilde{a}_{i,s} &< \frac{2\overline{P}_{i,s}}{\underline{b}_{b}-\underline{b}_{s}} \, \Leftrightarrow \, a_{i,s} > \frac{\underline{b}_{b}-\underline{b}_{s}}{2\overline{P}_{i,s}} \\
\label{a-cond-sb-1}
\frac{\overline{b}_{s}-\underline{b}_{s}}{\underline{b}_{b}-\overline{b}_{s}} &< \frac{\sum_{i \in \tV_{b}} \tilde{a}_{i,b}}{\sum_{i \in \tV_{s}} \tilde{a}_{i,s}} < \frac{\underline{b}_{b}-\overline{b}_{s}}{\overline{b}_{b}-\underline{b}_{b}}
\end{align}
\end{subequations}
\end{thm}

\begin{IEEEproof}
See Appendix \ref{apdix-1}.
\end{IEEEproof}

It can be observed from \eqref{ab-1} that in order to satisfying the global condition \eqref{a-cond-sb-1} private parameters $\underline{b}_{b}, \overline{b}_{b}, \underline{b}_{s}, \overline{b}_{s}$ need to be exchanged between prosumers. This is not acceptable from the privacy-preserving viewpoint of prosumers. Therefore, in the following other sufficient conditions are proposed, which are more conservative than that in \eqref{ab-1}, but are better in term of privacy protection. 

\begin{thm}
\label{para-cond-2}
Let $k > 3$, $0 < k_s < 2$, and $0 < k_b < 2$ such that 
\begin{equation}
\label{kkskb}
\frac{2}{k_b(k-2)} \, < \, \frac{-\sum_{i \in \tV_{b}} \underline{P}_{i,b}}{\sum_{i \in \tV_{s}} \overline{P}_{i,s}} \, < \, \frac{k_s(k-2)}{2}  
\end{equation}
Then the following conditions are sufficient to guarantee the strict feasibility of the constraints \eqref{bounded-power-p2p}--\eqref{pos-neg},
\begin{subequations}
\label{ab-2}
\begin{align}
\label{bb-selection-2}
\underline{\lambda}\, &\leq \,b_{i,s}\, < \,\underline{\lambda} + \frac{\overline{\lambda}-\underline{\lambda}}{k}  \\  
\label{bs-selection-2}
\underline{\lambda} + \frac{(k-1)(\overline{\lambda}-\underline{\lambda})}{k}\, &< \,b_{i,b}\, \leq \,\overline{\lambda}  \\
\label{a-cond-b-2}
\frac{\overline{\lambda}-\underline{\lambda}}{2\overline{P}_{i,s}} \, &< \, a_{i,s} \, \leq \, \frac{\overline{\lambda}-\underline{\lambda}}{k_s\overline{P}_{i,s}} \\
\label{a-cond-s-2}
\frac{\overline{\lambda}-\underline{\lambda}}{-2\underline{P}_{i,b}} \, &< \, a_{i,b} \, \leq \, \frac{\overline{\lambda}-\underline{\lambda}}{-k_b\underline{P}_{i,b}} 
\end{align}
\end{subequations}
\end{thm}

\begin{IEEEproof}
See Appendix \ref{apdix-2}.
\end{IEEEproof}

Conditions in \eqref{kkskb}--\eqref{ab-2} in fact show a way to satisfy conditions in \eqref{ab-1} of Theorem \ref{para-cond-1}, based on interval analysis. Once the new global condition \eqref{kkskb} is fulfilled, the parameters $a_{i,b}, a_{i,s}, b_{i,b}, b_{i,s}$ of prosumers' cost functions are selected in a fully decentralized manner as in \eqref{ab-2}. Additionally, only the upper or lower bounds on trading powers of prosumers are exchanged in \eqref{kkskb}, instead of private cost function parameters exchange in \eqref{a-cond-sb-1}. Thus, the privacy of prosumers are preserved. 

Note that there is a limitless number of choices for the parameters $k, k_s, k_b$ in \eqref{kkskb}, given the values of $\underline{P}_{i,b}$ and $\overline{P}_{i,s}$. Moreover, among those three parameters, $k$ is the most free parameter. Therefore, we can choose $k_s$ and $k_b$ first, then select $k$ appropriately to satisfy \eqref{kkskb}. One possibility is depicted in the following corollary. 
 
\begin{coro}
\label{coro-1}
Let $k_s=1$, $k_b=1$,  and choose $k>4$ such that
\begin{equation}
\label{kkskb-1}
\frac{2}{k-2} \, < \, \frac{-\sum_{i \in \tV_{b}} \underline{P}_{i,b}}{\sum_{i \in \tV_{s}} \overline{P}_{i,s}} \, < \, \frac{k-2}{2}  
\end{equation}
Then the strict feasibility of the constraints \eqref{bounded-power-p2p}--\eqref{pos-neg} is guaranteed by the selections of prosumers' cost function parameters in \eqref{ab-2}. 
\end{coro}

Corollary \ref{coro-1} is straightforwardly obtained from Theorem \ref{para-cond-2}, so a proof for it is not presented, for brevity. 

\begin{rem}
It is easy to select $k$ to satisfy \eqref{kkskb-1} or \eqref{kkskb}, simply by letting it as big as possible. As such, the interval $\left[\dfrac{2}{k-2}, \dfrac{k-2}{2}\right]$ is widened and will certainly contain $\dfrac{-\sum_{i \in \tV_{b}} \underline{P}_{i,b}}{\sum_{i \in \tV_{s}} \overline{P}_{i,s}}$ inside. The increase of $k$ also makes $b_{i,s}$ smaller and $b_{i,b}$ bigger, within the interval $\left[\underline{\lambda}, \overline{\lambda}\right]$, in order to satisfy the constraint \eqref{pos-neg}, i.e., to obtain successful trading. This is indeed in line with the heuristic learning strategy proposed in \cite{Nguyen-TPWRS20}, and can be utilized to analytically explain that strategy. 
\end{rem}

The global conditions \eqref{kkskb} and \eqref{kkskb-1} can be analytically verified in a decentralized fashion through prosumers cooperation. First, buying prosumers broadcast their lower bounds $\underline{P}_{i,b}$ of buying powers to all selling prosumers, and vice versa selling prosumers broadcast their upper bounds $\overline{P}_{i,s}$ of selling powers to all buying prosumers. Second, each buying prosumer calculates $\sum_{i \in \tV_{s}} \overline{P}_{i,s}$ and sends back to selling prosumers. Meanwhile, each selling prosumer computes $\sum_{i \in \tV_{b}} \underline{P}_{i,b}$ and sends back to buying prosumers. As the result, each prosumer can calculate the ratio $\frac{-\sum_{i \in \tV_{b}} \underline{P}_{i,b}}{\sum_{i \in \tV_{s}} \overline{P}_{i,s}}$. Denote this ratio by $\xi$. Consequently, prosumers choose $k, k_s, k_b$ to satisfy \eqref{kkskb} or \eqref{kkskb-1}. For example, it can be easily deduced that \eqref{kkskb-1} is equivalent to
\begin{equation}
	\label{kkskb-2}
	k > 2 + \max \left\{ \frac{2}{\xi}, 2\xi \right\}
\end{equation}
As such, each prosumer can choose an initial value of $k$ to satisfy \eqref{kkskb-2}, denoted by $k_i$. Afterward, a decentralized consensus algorithm is run by all prosumers to derive the average of $k_i$ which certainly satisfies \eqref{kkskb-2}. This average value is then utilized as $k$ by all prosumers to choose their private parameters as in \eqref{ab-2}.  

\begin{rem}
The results shown in Theorems \ref{para-cond-1}--\ref{para-cond-2} and Corollary \ref{coro-1} are derived for P2P energy systems with multiple buying and multiple selling prosumers. Those results will be simpler when only one selling or buying prosumer exists. 

If there is only one selling prosumer, then the first equality on the left hand side of \eqref{a-cond-sb-1} becomes trivial, because in this case $\overline{b}_{s}=\underline{b}_{s}$, hence it can be omitted. Similarly, if there is only one buying prosumer, then the second inequality on the right hand side of \eqref{a-cond-sb-1} is not needed. Accordingly, the results in Theorem \ref{para-cond-2} can be made simpler, as shown below.  
\end{rem}

\begin{coro}
\label{coro-2}
If there is only one buying prosumer, let $k>1$. Then parameters of prosumers' cost functions can be chosen as follows, 
\begin{subequations}
\label{ab-3}
\begin{align}
\label{b-selection-3}
\underline{\lambda}\, &\leq \,b_{i,s}\, < \,\underline{\lambda} + \frac{\overline{\lambda}-\underline{\lambda}}{k} \, < \,b_{b}\, \leq \,\overline{\lambda}  \\
\label{a-cond-b-3}
\frac{\overline{\lambda}-\underline{\lambda}}{2\overline{P}_{i,s}} \, &< \, a_{i,s}  \\
\label{a-cond-s-3}
\frac{\overline{\lambda}-\underline{\lambda}}{-2\underline{P}_{b}} \, &< \, a_{b} \, \leq \, \frac{kb_{b} - (k-1)\underline{\lambda} - \overline{\lambda}}{2\sum_{i \in \tV_{s}} \overline{P}_{i,s}} 
\end{align}
\end{subequations}
which are sufficient for the strict feasibility of the constraints \eqref{bounded-power-p2p}--\eqref{pos-neg}. Here, $a_b,b_b$ are parameters of the buying prosumer's cost function, and $\underline{P}_{b}$ is the minimum amount of power it wants to trade. 
\end{coro}

\begin{IEEEproof}
See Appendix \ref{apdix-3}.
\end{IEEEproof}

\begin{coro}
\label{coro-3}
If there is only one selling prosumer, let $k>1$. Then parameters of prosumers' cost functions can be chosen as follows, 
\begin{subequations}
\label{ab-4}
\begin{align}
\label{b-selection-4}
\underline{\lambda}\, &\leq \,b_{s}\, < \,\underline{\lambda} + \frac{\overline{\lambda}-\underline{\lambda}}{k} \, < \,b_{i,b}\, \leq \,\overline{\lambda}  \\
\label{a-cond-b-4}
\frac{\overline{\lambda}-\underline{\lambda}}{2\overline{P}_{s}} \, &< \, a_{s} \, \leq \, \frac{(k-1)\underline{\lambda} + \overline{\lambda} - kb_{s}}{-2(k-1)\sum_{i \in \tV_{s}} \underline{P}_{i,b}} \\
\label{a-cond-s-4}
\frac{\overline{\lambda}-\underline{\lambda}}{-2\underline{P}_{i,b}} \, &< \, a_{i,b} 
\end{align}
\end{subequations}
which are sufficient for the strict feasibility of the constraints \eqref{bounded-power-p2p}--\eqref{pos-neg}. Here, $a_{s},b_{s}$ are parameters of the selling prosumer's cost function, and $\overline{P}_{s}$ is the maximum amount of power it wants to trade. 
\end{coro}

\begin{IEEEproof}
%See Appendix \ref{apdix-4}.
The proof of this corollary is similar to that for Corollary \ref{coro-2}, hence is omitted here for brevity. 
\end{IEEEproof}

The results provided in Corollaries \ref{coro-2}--\ref{coro-3} include that in \cite{Nguyen-Access20} as a special case, where \cite{Nguyen-Access20} set $k=2$ and studied a specific application of electric vehicle wireless charging and discharging.

%------------------------------------------------------------------
\subsection{Privacy-Preserving P2P Energy Trading Mechanism}

To achieve the market clearing energy price  \eqref{eq-price} and power amounts \eqref{eq-power}, different optimization methods could be used to solve the forward optimization problem \eqref{p2p}, e.g., ADMM \cite{KZhang20,Baroche19,Baroche19conf,MorstynP2P19a,Nguyen-TPWRS20}, or primal-dual algorithms, relaxed consensus + innovation (RCI) \cite{Sorin19,Khorasany19}. However, the computational complexity will be accordingly increased due to the an additional iterative loop for executing such optimization algorithms. 
Therefore, privacy-preserving consensus algorithms will be employed to directly compute \eqref{eq-price}, while avoid exposing private parameters of prosumers, similar to that was used in \cite{Nguyen-Access20}. Each prosumer/agent $i, i=1,\ldots,n$ creates a vector $x_i \in \bR^{2}$ with initial value:
\begin{equation}
	\label{mas-init}
	\scalebox{1}{$
	\begin{aligned}
		x_{i}(0) = \left[\frac{b_{i,b}}{a_{i,b}},\frac{1}{a_{i,b}}\right]^T, i \in \tV_b; 
		x_{i}(0) = \left[\frac{b_{i,s}}{a_{i,s}},\frac{1}{a_{i,s}}\right]^T, i \in \tV_s 
	\end{aligned}
	$}	
\end{equation}
Subsequently, at each time step $k \geq 0$, each prosumer/agent creates random noises $w_{i,1}(k)$ and $w_{i,2}(k)$ defined by:
\begin{equation}
	\label{mask}
	w_{i,\ell}(k) = \left\{ 
	\begin{aligned}
		& \zeta_{i,\ell}(0), \quad & {\rm if} \; k=0 \\
		& \alpha_{i}^{k}\zeta_{i,\ell}(k)-\alpha_{i}^{k-1}\zeta_{i,\ell}(k-1), \quad & {\rm otherwise}
	\end{aligned}
	\right.
\end{equation}
for $\ell=1,2$, where $\zeta_{i,\ell}(k)$ are independently generated Gaussian random variables from a standard normal distribution, and $0<\alpha_{i}<1$ are constants. Employing those random noises, each peer/prosumer obtains a masked state vector
\begin{equation}
	\label{masked-state}
	\tilde{x}_i(k) = x_i(k) + [w_{i,1}(k),w_{i,2}(k)]^T
\end{equation} 
Then each prosumer runs the secure consensus algorithm,
\begin{equation}
	\label{secure-consensus}
	x_{i}(k+1) = a_{ii}\tilde{x}_{i}(k) + \sum_{j \in \mathcal{N}_{i}}{a_{ij}\tilde{x}_{i}(k)},i=0,1,\ldots,n
\end{equation}
The weights $a_{ij}$ can be determined by different ways (see e.g., \cite{Nguyen-TSG17}, \cite{Nguyen-TIE19}). 
It can be proved similarly to \cite{YMo17} that the average consensus is asymptotically achieved for all peers, i.e. $\displaystyle \lim_{k \rightarrow \infty} x_i(k) = x_{{\rm ave}}$, where $x_{{\rm ave}} = [x_{{\rm ave},1},x_{{\rm ave},2}]^T$, and $x_{{\rm ave},1}, x_{{\rm ave},2}$ converge precisely to $\frac{\sum_{i=1}^{n} b_i/a_i}{n}, \frac{\sum_{i=1}^{n}1/a_i}{n}$, respectively. 
Thus, the P2P market clearing energy price can be computed by 
\begin{equation}
\label{eq-price-masked}
\lambda^\ast = \frac{x_{{\rm ave},1}}{x_{{\rm ave},2}}
\end{equation}
 
Finally, the proposed cooperative learning approach for obtaining successful and desired energy transactions for prosumers/agents is summarized in Algorithm \ref{ms-mb-learning}. %, and the secured consensus algorithm above is also included for computing the P2P energy market clearing price.

\begin{algorithm}
	\caption{Multi-Seller-Multi-Buyer Cooperative Learning for P2P Energy Trading}
\begin{algorithmic}

	%\SetKwInOut{Input}{Input}
	%\SetKwInOut{Output}{Output}
	
	\STATE Buying and selling prosumers set their own ranges of energy prices $[\underline{\lambda}_i(0),\overline{\lambda}_i(0)]$, and energy amounts $[\underline{P}_{i,b},0)$, $(0,\overline{P}_{i,s}]$ for trading; 
			
	\% {\it Step 1: Negotiation of a common energy price range};

	\FOR {$1 \leq k \leq max\_iter$}
	
	\STATE Prosumers run the consensus algorithm \eqref{secure-consensus} with normal states $x_i(k)$ and without random noises $w_{i,1}, w_{i,2}$;
	
	\IF {$k=max\_iter$, or $|\underline{\lambda}_i(k+1)-\underline{\lambda}_i(k)| \leq \epsilon$, $|\overline{\lambda}_i(k+1)-\overline{\lambda}_i(k)| \leq \epsilon \; \forall \; i=1,\ldots,n,$}
			\STATE break;
	\ENDIF			
	\ENDFOR
			
	\STATE Prosumers obtain the common P2P energy trading price range $[\underline{\lambda},\overline{\lambda}]$;	

	\% {\it Step 2: Selection of cost function parameters}
	
	\STATE Buying prosumers choose parameters as in \eqref{bs-selection-2} and \eqref{a-cond-s-2}; 
	
	\STATE Selling prosumers choose parameters as in \eqref{bb-selection-2} and \eqref{a-cond-b-2}; 
	
	\% {\it Step 3: Negotiation of P2P trading price and amounts}	
	
	\FOR {$1 \leq k \leq max\_iter$}
	\STATE Prosumers run the masked consensus algorithm \eqref{secure-consensus};
	
	\IF {$k=max\_iter$, or $\|\tilde{x}_i(k+1)-\tilde{x}_i(k)\|_2 \leq \epsilon \; \forall \; i=1,\ldots,n,$}
			\STATE break;
	\ENDIF		
	\ENDFOR	
		
	\STATE Prosumers calculate the P2P energy market clearing price $\lambda^\ast$ by \eqref{eq-price-masked}, and their energy trading amounts by \eqref{eq-power};	
	
\end{algorithmic}
	\label{ms-mb-learning}
\end{algorithm}

%%%%%%%%%%%%%%%%%%%%%%%%%%%%%%%%%%%%%%%%%%%%%%%%%%%%%%%%%%%%%%%%%%%%%%%%%%%%%%%%%%%%%%%%%%%%%%%%%%%%%%%%%%%%%%%%%%%%
\section{Case Study}
\label{num}

\begin{figure}[htpb!]
		\centering
		\includegraphics[scale=0.35]{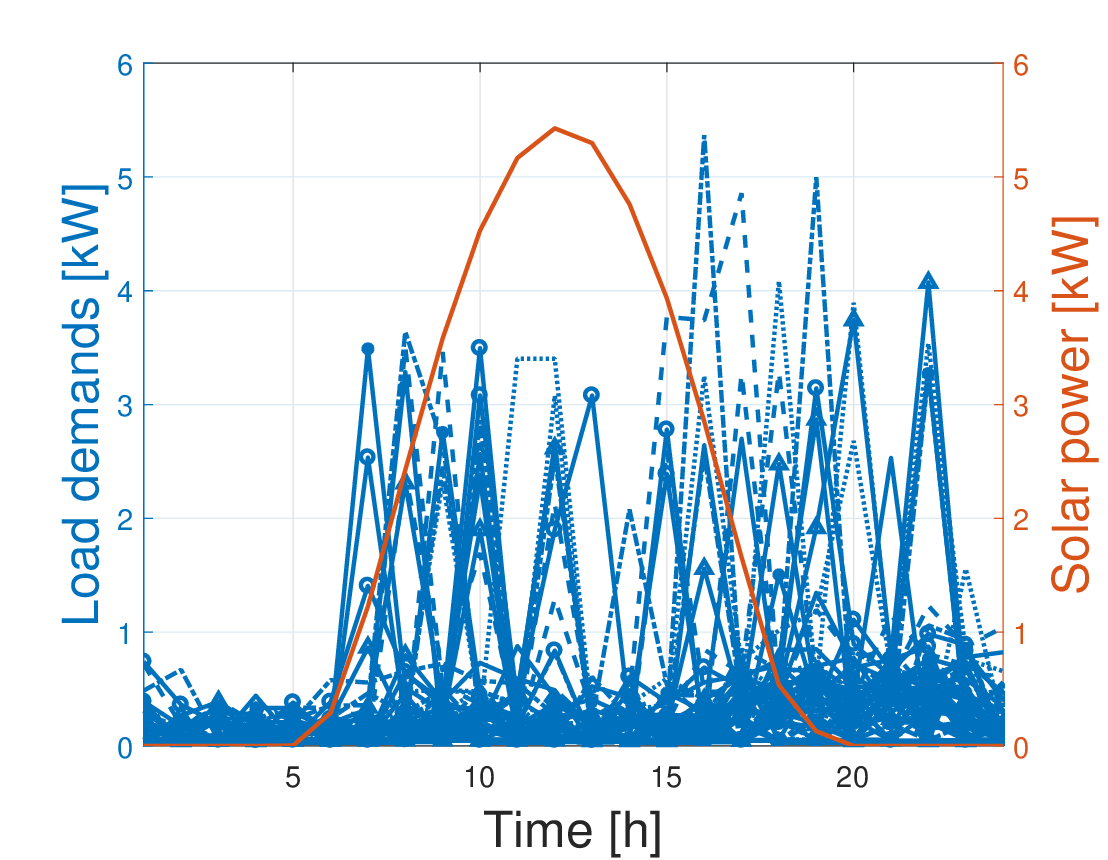}
		%\vspace{-15mm}
		\caption{Load demands and assumed solar generation in the IEEE European Low Voltage Test Feeder.}
		\label{eu_1h_loads_solar}
	\end{figure}

This section is intended to illustrate the proposed cooperative learning approach between prosumers/agents by applying to the IEEE European Low Voltage Test Feeder consisting of 55 nodes \cite{eu_55}. Similarly to \cite{Nguyen-TPWRS20}, we assume here the existence of 5.5kW rooftop solar power generation for 25 nodes and 3kWh battery systems for the other 30 nodes, hence each node is a potential prosumer who can perform P2P energy trading every hour. Realistic load patterns of all nodes, displayed in Figure \ref{eu_1h_loads_solar} at each hour during a day, are provided by \cite{eu_55}. Solar power generation is computed based on the average daily global solar irradiance data given in \cite{eu_solar_data} for Spain in July, similarly to that in \cite{Nguyen-TPWRS20}. As such, 25 nodes with rooftop solar power generation have approximately 2kW of power for selling at noon, whereas the other 30 nodes can buy maximum 3kW of power, i.e. $\overline{P}_{i,s}=2$ and $\underline{P}_{i,b}=-3$. 

There was a feed-in-tariff system for renewable energy in Spain but it was terminated in 2014, and the current electricity price in Spain is about 0.236 USD/kWh, i.e. about 24.8 JPY/kWh. Hence, we assume here that selling prosumers randomly set their intervals of preferred energy price within $[20, 23.8]$ JPY/kWh, whereas buying prosumers randomly select their expected energy price within $[19, 23]$ JPY/kWh. 

\begin{figure}[htpb!]
		\centering
		\includegraphics[scale=0.35]{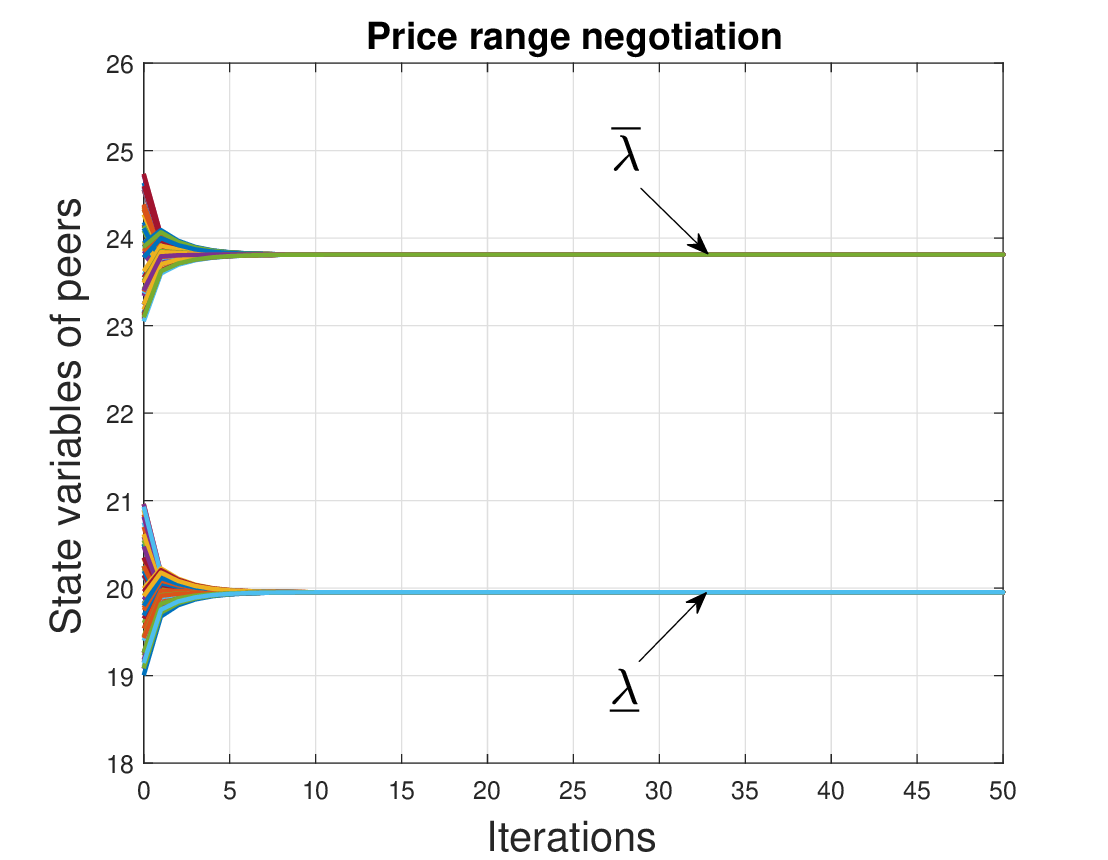}
		%\vspace{-15mm}
		\caption{Negotiation of P2P energy trading price range.}	
		\label{eu_price_range}
	\end{figure}

Consequently, following Algorithm \ref{ms-mb-learning}, prosumers first negotiate to obtain an agreed price interval by running the consensus algorithm \eqref{secure-consensus} without added random noises. The negotiation results are shown in Figure \ref{eu_price_range}, which reveal that $\overline{\lambda}=23.81$ JPY/kWh and $\underline{\lambda}=19.95$ JPY/kWh. Next, prosumers cooperative learn to check the global condition \eqref{kkskb}, or \eqref{kkskb-2} for simplicity. It then turns out that the global parameter $k$ should satisfy $k>5.6$. As discussed after \eqref{kkskb-2}, prosumers can initially choose their local copies of $k$ to fulfill \eqref{kkskb-2}, and then run a consensus algorithm to derive a common, global value of $k$. Here, we assume, for the sake of conciseness, that all prosumers reach a consensus on $k$ to be 5.7. Subsequently, all prosumers follow step 2 in Algorithm \ref{ms-mb-learning} to locally and randomly choose their cost function parameters in the associated intervals specified in \eqref{ab-2}. 

\begin{figure}[htpb!]
		\centering
		\includegraphics[scale=0.35]{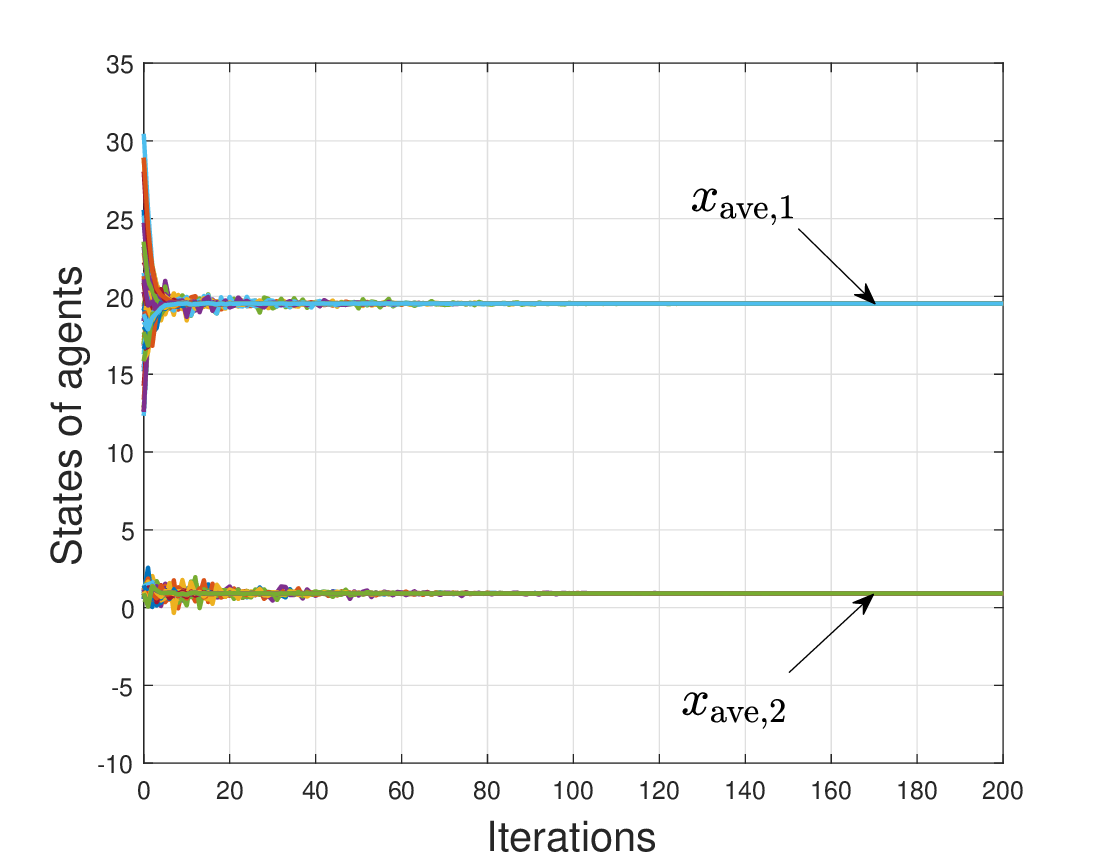}
		%\vspace{-15mm}
		\caption{Consensus of masked states of peers/agents.}	
		\label{eu_masked_css}
	\end{figure}

\begin{figure}[htpb!]
		\centering
		\includegraphics[scale=0.35]{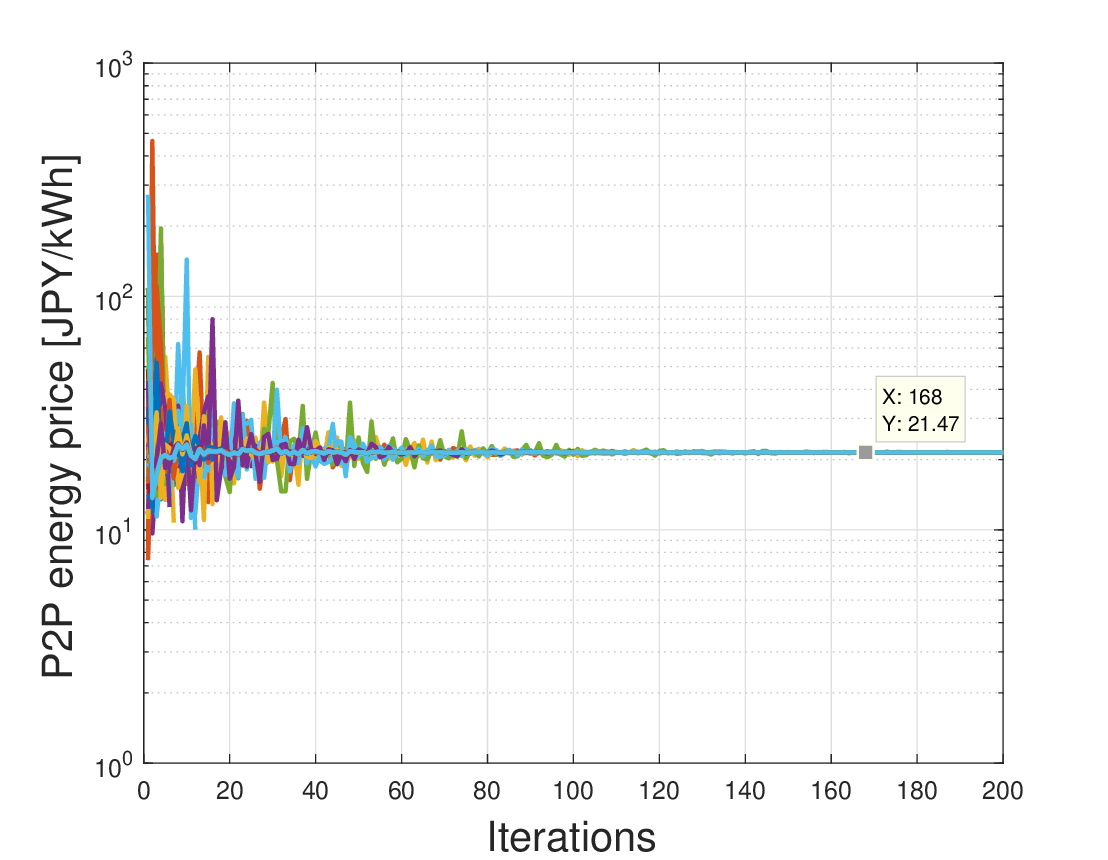}
		%\vspace{-15mm}
		\caption{Consensus of masked price negotiation.}	
		\label{eu_masked_price}
	\end{figure}

Finally, prosumers execute the masked consensus algorithm \eqref{secure-consensus} with added Gaussian noises to protect their data privacy while achieving a consensus on the P2P energy trading price. States of prosumers/agents and the P2P market-clearing price are then depicted in Figures \ref{eu_masked_css}--\ref{eu_masked_price}. As observed in Figure \ref{eu_masked_price}, the P2P market-clearing price is 21.47 JPY/kWh which indeed lies between $\underline{\lambda}=19.95$ JPY/kWh and $\overline{\lambda}=23.81$ JPY/kWh.  

\begin{figure}[htpb!]
		\centering
		\includegraphics[scale=0.35]{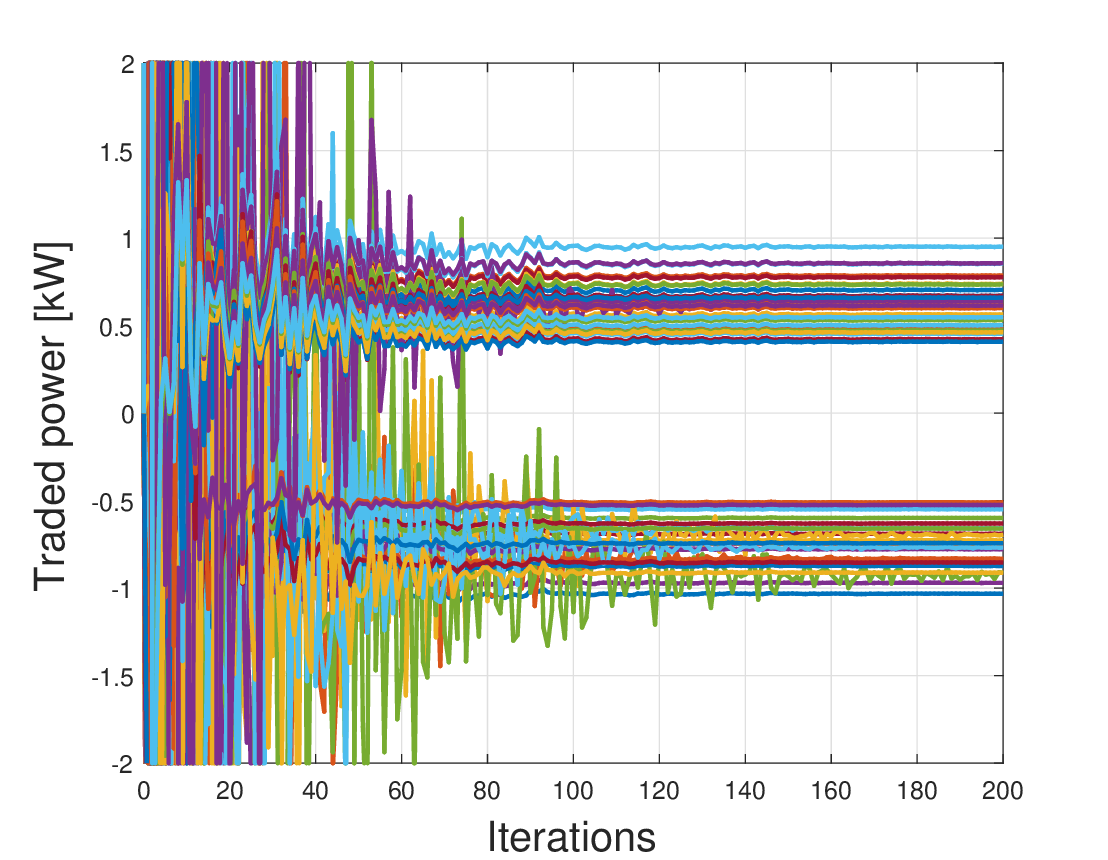}
		%\vspace{-15mm}
		\caption{Total P2P trading power of peers/agents.}	
		\label{eu_masked_energy}
	\end{figure}

The obtained P2P power trading between prosumers are exhibited in Figure \ref{eu_masked_energy}, which are all successful and within their limits. Thus, all simulation results confirm the effectiveness of the proposed cooperative learning apparoach for prosumers. 
			
\begin{figure}[htpb!]
		\centering
		\includegraphics[scale=0.35]{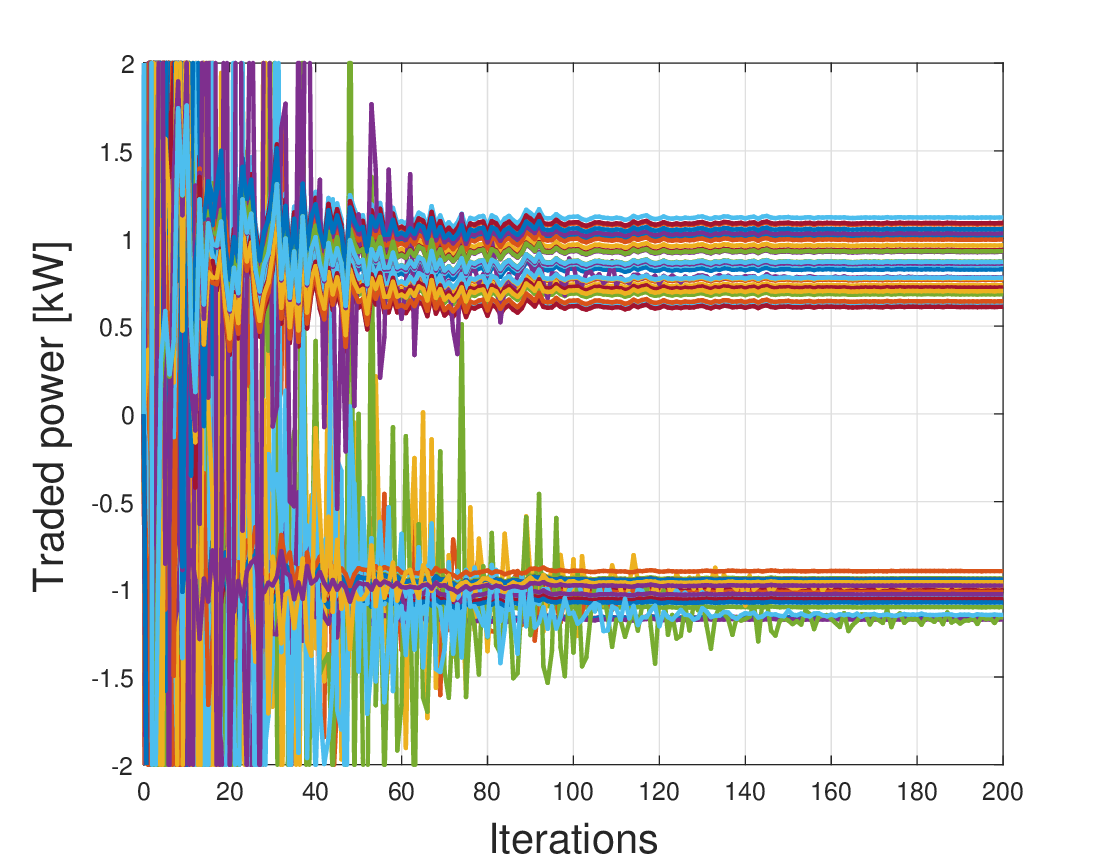}
		%\vspace{-15mm}
		\caption{Increased total P2P trading power of peers/agents after learning.}	
		\label{eu_masked_energy_new}
	\end{figure}

The trading power amount of prosumers can be increased by using the heuristic learning strategy in \cite{Nguyen-TPWRS20}, as depicted in Figure \ref{eu_masked_energy_new}, when the distances between current values of $a_{i,s}, a_{i,b}$ to their lower bounds are decreased 16 times. However, trading power amounts of prosumers cannot be arbitrarily increased, because $a_{i,s}$ and $a_{i,b}$ cannot be arbitrarily decreased but are lower bounded (see \eqref{a-cond-b-2}--\eqref{a-cond-s-2}). 
Compared to simulation results for the same system in \cite{Nguyen-TPWRS20}, the trading amounts here are smaller because of different values of $a_{i,s}, a_{i,b}$ and $b_{i,s}, b_{i,b}$, however their selections here are systematic and analytical, while that in \cite{Nguyen-TPWRS20} were heuristic. 
	
\section{Conclusion}
\label{sum}

A cooperative learning approach is proposed in this papers for prosumers/agents participating in P2P energy systems to analytically select their cost function parameters to achieve successful energy transactions with desired price and amounts. Inverse optimization is used to mathematically formulate the problem, and interval analysis is employed to solve it. A few global inequalities are subject to be learned by prosumers, which are dependent on the lower and upper bounds of their expected energy trading amounts. Then prosumers can locally choose their cost function parameters to assure the final negotiated energy price and trading amounts belong to their desired intervals. 
%Moreover, the inter-prosumer negotiation can be secured using existing strategies, e.g. privacy-preserving consensus, or cryptography. 
Numerical simulations from a case study shows the effectiveness of the proposed approach.    

The next research should develop a new cooperative learning approach to cover also the selection of trade weights on energy transactions, which is challenging because energy prices are no longer unique (see e.g.,\cite{Nguyen-TPWRS20} for more details).   

%In the next research, the proposed cooperative learning approach should be developed to cope with the selection of trade weights to reflect more energy preferences from prosumers. This would be much more complex than the situation in the current research, where energy price for each pair of prosumers would be different whose calculation is also more complicated (see e.g.,\cite{Nguyen-TPWRS20} for the computation of such prices).   

%%%%%%%%%%%%%%%%%%%%%%%%%%%%%%%%%%%%%%%%%%%%%%%%%%%%%%%%%%%%%%%%%%%%%%%%%%%%%%%%%%%%%%%%%%%%%%%%%%%%%%%%%%%%%%%%%%%%

\section*{Acknowledgement}
This research was financially supported by JSPS Kakenhi Grant Number JP19K15013.

\section{Appendix}

%-------------------------------------------------------
\subsection{Proof of Theorem \ref{para-cond-1}}
\label{apdix-1}

First, to guarantee the successful trading of prosumers, the constraint \eqref{pos-neg} must be satisfied. From \eqref{eq-power}, this means $\lambda^\ast < b_{i,b}$ for buying prosumers, and $\lambda^\ast > b_{i,s}$ for selling prosumers. We have
\begin{align}
\label{thm1-pf-eq1}
\lambda^\ast - b_{i,b} &= \sum_{i \in \tV_{b}} \frac{b_{j,s}-b_{i,b}}{a_{i,b}} + \sum_{i \in \tV_{s}} \frac{b_{j,b}-b_{i,b}}{a_{i,s}} \notag \\
&\leq \sum_{i \in \tV_{b}} \frac{\overline{b}_{b}-\underline{b}_{b}}{a_{i,b}} + \sum_{i \in \tV_{s}} \frac{\overline{b}_{s}-\underline{b}_{b}}{a_{i,s}}
\end{align}
Hence, a sufficient condition for $\lambda^\ast < b_{i,b}$ is that the right hand side of \eqref{thm1-pf-eq1} is negative, which is equivalent to
\begin{align}
\label{thm1-pf-eq2}
\sum_{i \in \tV_{b}} \tilde{a}_{i,b}(\overline{b}_{b}-\underline{b}_{b}) < \sum_{i \in \tV_{s}} \tilde{a}_{i,s}(\underline{b}_{b}-\overline{b}_{s})
\end{align}
On the other hand,
\begin{align}
\label{thm1-pf-eq3}
\lambda^\ast - b_{i,s} &= \sum_{i \in \tV_{b}} \frac{b_{j,s}-b_{i,s}}{a_{i,b}} + \sum_{i \in \tV_{s}} \frac{b_{j,b}-b_{i,s}}{a_{i,s}} \notag \\
&\geq \sum_{i \in \tV_{b}} \frac{\underline{b}_{b}-\overline{b}_{s}}{a_{i,b}} + \sum_{i \in \tV_{s}} \frac{\underline{b}_{s}-\overline{b}_{s}}{a_{i,s}}
\end{align}
Therefore, a sufficient condition for $\lambda^\ast > b_{i,s}$ is that the right hand side of \eqref{thm1-pf-eq3} is positive, which is equivalent to
\begin{align}
\label{thm1-pf-eq4}
\sum_{i \in \tV_{b}} \tilde{a}_{i,b}(\underline{b}_{b}-\overline{b}_{s}) > \sum_{i \in \tV_{s}} \tilde{a}_{i,s}(\overline{b}_{s}-\underline{b}_{s})
\end{align}
With the parameters $b_{i,b},b_{i,s}$ selected as in \eqref{b-selection-1} and the numbers of selling and buying prosumers are more than one, the inequalities \eqref{thm1-pf-eq2} and \eqref{thm1-pf-eq4} lead to \eqref{a-cond-sb-1}.

Next, the constraint \eqref{bounded-power-p2p} is equivalent to
\begin{equation}
\label{thm1-pf-eq5}
\begin{aligned}
\frac{\tilde{a}_{i,b}}{2}(b_{i,b}-\lambda^\ast) &\leq -\underline{P}_{i,b} \\
\frac{\tilde{a}_{i,s}}{2}(\lambda^\ast-b_{i,s}) &\leq \overline{P}_{i,s}
\end{aligned}
\end{equation}
It can be deduced that 
\begin{align}
\label{thm1-pf-eq6}
\frac{\tilde{a}_{i,b}}{2}(b_{i,b}-\lambda^\ast) &\leq \frac{\tilde{a}_{i,b}}{2}(b_{i,b}-\overline{b}_{s}) \leq \frac{\tilde{a}_{i,b}}{2}(\overline{b}_{b}-\overline{b}_{s})
\end{align} 
Consequently, the following condition is sufficient for the first inequality in \eqref{thm1-pf-eq5},
\begin{align}
\label{thm1-pf-eq7}
\frac{\tilde{a}_{i,b}}{2}(\overline{b}_{b}-\overline{b}_{s}) < -\underline{P}_{i,b}
\end{align}
which is exactly \eqref{a-cond-s-1}. 
On the other hand,
\begin{align}
\label{thm1-pf-eq8}
\frac{\tilde{a}_{i,s}}{2}(\lambda^\ast-b_{i,s}) &\leq \frac{\tilde{a}_{i,s}}{2}(\underline{b}_{b}-b_{i,s})  \leq \frac{\tilde{a}_{i,s}}{2}(\underline{b}_{b}-\underline{b}_{s})
\end{align} 
Thus, the following condition is sufficient for the second inequality in \eqref{thm1-pf-eq5},
\begin{align}
\label{thm1-pf-eq9}
\frac{\tilde{a}_{i,s}}{2}(\underline{b}_{b}-\underline{b}_{s}) < \overline{P}_{i,s}
\end{align} 
which is the same as \eqref{a-cond-b-1}.

%-------------------------------------------------------
\subsection{Proof of Theorem \ref{para-cond-2}}
\label{apdix-2}

Obviously, the conditions in \eqref{bb-selection-2}--\eqref{bs-selection-2} is a way to satisfy \eqref{b-selection-1}, where $b_{i,s}$ and $b_{i,b}$ are forced to lie in smaller intervals which are disjointed. Moreover, we can easily deduce that 
\begin{equation}
\label{thm2-pf-eq1}
\begin{aligned}
\frac{\overline{b}_{s}-\underline{b}_{s}}{\underline{b}_{b}-\overline{b}_{s}} &< \frac{1/k}{(k-2)/k} = \frac{1}{k-2} \\
 \frac{\underline{b}_{b}-\overline{b}_{s}}{\overline{b}_{b}-\underline{b}_{b}} &> \frac{(k-2)/k}{1/k} = k-2
\end{aligned} 
\end{equation}
On the other hand, the selections of $a_{i,s}$ and $a_{i,b}$ in \eqref{a-cond-b-2}--\eqref{a-cond-s-2} lead to
\begin{align}
\label{thm2-pf-eq2}
\frac{-k_b\sum_{i \in \tV_{b}} \underline{P}_{i,b}}{2\sum_{i \in \tV_{s}} \overline{P}_{i,s}} &< \frac{\sum_{i \in \tV_{b}} \tilde{a}_{i,b}}{\sum_{i \in \tV_{s}} \tilde{a}_{i,s}} < \frac{-2\sum_{i \in \tV_{b}} \underline{P}_{i,b}}{k_s\sum_{i \in \tV_{s}} \overline{P}_{i,s}}
\end{align} 
Subsequently, substituting \eqref{kkskb} into \eqref{thm2-pf-eq2} gives us
\begin{align}
\frac{1}{k-2} < \frac{\sum_{i \in \tV_{b}} \tilde{a}_{i,b}}{\sum_{i \in \tV_{s}} \tilde{a}_{i,s}} < k-2
\end{align} 
which, together with \eqref{thm2-pf-eq1}, clearly shows that the condition \eqref{a-cond-sb-1} is satisfied.

%-------------------------------------------------------
\subsection{Proof of Corollary \ref{coro-2}}
\label{apdix-3}

If there is only one buying prosumer, then the right hand side of \eqref{a-cond-sb-1} is not needed. In addition, \eqref{b-selection-3}, \eqref{a-cond-b-3}, and \eqref{a-cond-s-3} trivially guarantees \eqref{b-selection-1}, \eqref{a-cond-b-1}, and \eqref{a-cond-s-1}, respectively. 

With the choices of $b_s$ and $b_{i,s}$ as in \eqref{b-selection-3}, we obtain
\begin{equation}
\label{coro2-pf-eq1}
\frac{\overline{b}_{s}-\underline{b}_{s}}{b_{b}-\overline{b}_{s}} < \frac{(\overline{\lambda} - \underline{\lambda})/k}{b_{b} - \underline{\lambda} - \frac{\overline{\lambda}-\underline{\lambda}}{k}} = \frac{\overline{\lambda} - \underline{\lambda}}{kb_{b} - (k-1)\underline{\lambda} - \overline{\lambda}} 
\end{equation}
Next, the right hand side of \eqref{a-cond-s-3} and \eqref{a-cond-b-3} lead to
\begin{align}
\label{coro2-pf-eq2}
\frac{\tilde{a}_{b}}{\sum_{i \in \tV_{s}} \tilde{a}_{i,s}} > \frac{\overline{\lambda} - \underline{\lambda}}{kb_{b} - (k-1)\underline{\lambda} - \overline{\lambda}} 
\end{align} 
The combination of \eqref{coro2-pf-eq1} and \eqref{coro2-pf-eq2} gives us the left hand side of \eqref{a-cond-sb-1}. 

%%-------------------------------------------------------
%\subsection{Proof of Corollary \ref{coro-3}}
%\label{apdix-4}

%%The proof of this Corollary is similar to that for Corollary \ref{coro-2}, hence it is omitted here for brevity. 

%If there is only one selling prosumer, then the left hand side of \eqref{a-cond-sb-1} is not needed. Moreover, the conditions \eqref{b-selection-1}, \eqref{a-cond-b-1}, and \eqref{a-cond-s-1} are satisfied by \eqref{b-selection-4}, \eqref{a-cond-b-4}, and \eqref{a-cond-s-4}, respectively. 
%
%Then the choices of $b_b$ and $b_{i,b}$ as in \eqref{b-selection-4} give us
%\begin{equation}
%\label{coro3-pf-eq1}
%\frac{\underline{b}_{b}-b_{s}}{\overline{b}_{b}-\underline{b}_{b}} > \frac{\underline{\lambda} + \frac{\overline{\lambda}-\underline{\lambda}}{k}  - b_{s}}{(k-1)(\overline{\lambda} - \underline{\lambda})/k} = \frac{(k-1)\underline{\lambda} + \overline{\lambda} - kb_{s}}{(k-1)(\overline{\lambda} - \underline{\lambda})} 
%\end{equation}
%On the other hand, it is deduced from the right hand side of \eqref{a-cond-b-4} and \eqref{a-cond-s-4} that
%\begin{align}
%\label{coro3-pf-eq2}
%\frac{\sum_{i \in \tV_{s}} \tilde{a}_{i,b}}{\tilde{a}_{s}} < \frac{(k-1)\underline{\lambda} + \overline{\lambda} - kb_{s}}{(k-1)(\overline{\lambda} - \underline{\lambda})} 
%\end{align} 
%Hence, the right hand side of \eqref{a-cond-sb-1} is guaranteed from \eqref{coro3-pf-eq1} and \eqref{coro3-pf-eq2}. 

%%%%%%%%%%%%%%%%%%%%%%%%%%%%%%%%%%%%%%%%%%%%%%%%%%%%%%%%%%%%%%%%%%%%%%%%%%%%%%%%%%%%%%%%%%%%%%%%%%%%%%%%%%%%%%%%%%%
\bibliographystyle{plain}
\bibliography{dp}
%\vspace{-1.2cm}

 %\begin{IEEEbiography}[{\includegraphics[width=1in,height=1.25in,clip,keepaspectratio]{Hoa}}]{Dinh Hoa Nguyen} (S'09 M'15) received a Ph.D. degree from The University of Tokyo in 2014. He is currently an Assistant Professor at Kyushu University, Fukuoka, Japan. His research is on modeling, optimization, and control toward low-carbon and autonomous energy systems, with particular focuses on renewables and distributed energy resources, smart grid, artificial intelligence, multi-agent systems, and decentralized optimization.
 %\end{IEEEbiography}

\end{document}